\documentstyle[epsfig,12pt]{article}
\newcommand{\beq}{\begin{equation}}
\newcommand{\eeq}{\end{equation}}
\newcommand{\bea}{\begin{eqnarray}}
\newcommand{\eea}{\end{eqnarray}}
\def\Dzero{D$\emptyset$}
\def\Re{{\rm Re}}
\def\cW{{\cal W}}

\begin{document}
\pagestyle{empty}
\begin{flushright}
{CERN-TH/96-225}\\
{DTP/96/78}\\
{LBNL-39238}\\
\end{flushright}
\vspace*{5mm}
\begin{center}
{\large {\bf Hadronic Antenna Patterns to Distinguish Production
Mechanisms for Large-$E_T$ Jets}}\\
\vspace*{1cm}
{\bf John Ellis$^*$}\\
\vspace{0.3cm}
Theoretical Physics Division, CERN \\
1211 Geneva 23, Switzerland\\
\vspace{0.3cm}
{\bf Valery A. Khoze} and {\bf W. James Stirling} \\
\vspace{0.3cm}
Department of Physics, University of Durham,\\
Durham DH1 3LE, United Kingdom\\
%\vspace*{2cm}
\vspace*{1.5cm}
{\bf ABSTRACT} \\ \end{center}
\vspace*{5mm}
\noindent
Hadronic antenna patterns provide a tool able to
diagnose different patterns of colour flow in
large-$E_T$ jet events. They reflect the underlying
short-distance dynamics, and are sensitive to colour
coherence and interference between the initial- and
final-state partons. We discuss how hadronic antenna
patterns may be used on large-$E_T$ events from
the Fermilab Tevatron or the CERN LHC to distinguish
between conventional QCD and new physics production mechanisms
such as a possible $Z'$ boson or compositeness.

%\vspace*{5cm}
\vfill

\begin{flushleft}
$^*$ This work was supported in part by the Director, Office of
Energy Research, Office of Basic Energy Science of the U.S.
Department of Energy, under Contract DE-AC03-76SF00098. \\
{~}\\
CERN-TH/96-225\\
DTP/96/78\\
LBNL-39238\\
August 1996
\end{flushleft}
%\vfill
\eject
%\pagestyle{empty}
%\clearpage\mbox{}\clearpage

\setcounter{page}{1}
\pagestyle{plain}

\section{Introduction}

Stimulated by recent data \cite{CDF,D0}
from the Fermilab Tevatron collider   and by the prospects
for future experiments with the CERN  LHC,
there is currently much interest in the possible
interpretations of hadronic jets at large transverse energy $E_T$
\cite{Zprime,susy}.   It is premature to 
focus on speculative interpretations of
the CDF data \cite{CDF} until they have been reconciled
satisfactorily with those from \Dzero\ \cite{D0}, and   until there
there is consensus whether it is possible
to make a new global fit to parton distributions
within the proton which accommodates the new data: for
contrasting approaches, see \cite{CTEQ,GMRS}.
Nevertheless, it is appropriate to ask   whether
there are other ways in which one could in
principle distinguish between a conventional QCD
interpretation of large-$E_T$ jet data and possible new
physics such as a $Z'$ boson \cite{Zprime} or compositeness
\cite{ELP}. This question is also important for future
experiments at the LHC, which may well be confronted by it
in their own large-$E_T$ data a decade or so hence.

The purpose of this paper is to publicize a diagnostic
tool which has been proposed in the literature but has not, to
our knowledge, been discussed in connection with current
large-$E_T$ data: it is the pattern of hadronic energy
flow around the large-$E_T$ jet axes.

It has been known for a long time \cite{DKT,DKMT} that
the structure of multi-jet events in hard processes is
influenced by the underlying colour dynamics at short distances.
Detailed features of the parton shower, in particular the
flow of colour quantum numbers, control the distributions
of colour-singlet hadrons in the final state \cite{book}.
To leading order in the large-$N_c$ limit, the analytic results
for such antenna patterns coincide with the Lund string picture \cite{Lund}.
The first, and still the best, example of such a colour-related
phenomenon is the so-called string \cite{string} or drag \cite{drag}
effect in $e^+ e^- \rightarrow \bar q q g$ events, which is very
well established experimentally, see for example Refs.~\cite{h,ms}.
 Colour coherence effects have
also been seen very clearly in multi-jet events in $\bar p p$
collisions \cite{exp}.

Patterns of hadron production in the large-$E_T$
event plane have been measured, and shown \cite{exp} to differ
significantly from the predictions of Monte Carlo models
that do not include colour coherence and interference
effects. On the other hand, the measurements agree very well
with the predictions of Monte Carlo models, such as HERWIG
\cite{HERWIG}, JETSET \cite{JETSET}, ARIADNE \cite{ARIADNE} and JETRAD
\cite{JETRAD},
which incorporate QCD colour coherence with interfering
gluons\footnote{There are interesting differences between Monte Carlo
results and the exact analytical QCD expectations \cite{DKTSJNP}, but these do not
concern us here.}. In particular,
initial/final-state colour interference effects have been seen
very clearly in the data. Thus the hadronic structure of a
multi-jet event draws its colour portrait, and can be
regarded as a ``partonometer" mapping the basic interaction
process.

A natural idea which has been proposed
%\cite{DKT,KStanford,PhysColl,DKS,DKTSJNP}, is to
\cite{DKT,DKTSJNP,KStanford,DKS,MarWeb,ZEPPEN,emw}, is to
use this ``partonometry' to help
distinguish new physics signals from backgrounds due to
conventional QCD. One example of particular interest to the
LHC is the possible use of rapidity gaps to favour Higgs
%production events \cite{PhysColl,DKS}, and the pattern
production events \cite{DKT,DKS}, and the pattern
of QCD radiation in $t \bar t $ production has been examined \cite{KSt}.
Our suggestion here is to use ``partonometry" to dissect
the colour structure of the large-$E_T$ jet events
observed by CDF \cite{CDF} and   \Dzero\ \cite{D0}, as a way to
distinguish between conventional QCD and possible new
production mechanisms such as a $Z'$ boson \cite{Zprime} or
compositeness \cite{CDF,ELP}.

For any given pattern of colour flow between the initial
and final states, there is a characteristic \cite{DKTSJNP,MarWeb,emw}
pattern of gluon emission and hence hadronic energy flow
in the transverse event plane. This is universal for
hadron energies in the range $\Lambda_{QCD} \ll E \ll \sum E_T$,
in the approximation of a large number of colours $N_c$. This
should be a good approximation for
secondary jet emission in the $ q q(q \bar q)$, $q g$ and
$gg$ scattering processes of most interest for the Tevatron
and the LHC. In this paper, we collect and compile the
relevant formulae for conventional QCD, $Z'$ and
composite-model contributions to the large-$E_T$ cross
section, and discuss how these may be used to distinguish
between these possible production mechanisms.

\section{An Example of the Diagnostic Power of Hadronic Antenna
Patterns}
\label{sec:sec2}

We begin by considering a simple illustrative example,
the  subprocess $q \bar q \to q' \bar q'$, 
where $q$ and $q'$ denote different quark flavours,
 which makes a distinctive
contribution to the hadronic energy  flow  in 
large-$E_T$ jet production at the Tevatron collider.
In QCD, this process is dominated by $s$-channel gluon exchange.
Suppose that there is an additional non-standard contribution
from the $s$-channel exchange of a new, heavy $Z'$ vector boson,
$q \bar q \to Z' \to q' \bar q'$. If we assume, for simplicity,
that the $Z'$ has a vector coupling of strength $g'$ to the quarks,
then the matrix element squared is
\begin{equation}
\overline{\sum}\vert{\cal M}\vert^2 \; =\;
{g_s^4 C_F \over N_c}\;  {u^2 + t^2 \over s^2}\; + \;
 2 g'^4\; { u^2 + t^2  \over (s-M_{Z'}^2)^2
+ M_{Z'}^2 \Gamma_{Z'}^2} \; ,
\end{equation}
where $C_F=(N_c^2-1)/2N_c$, and
$M_{Z'}$ and $\Gamma_{Z'}$ are the $Z'$ mass and width
respectively. If the final-state quarks are produced at wide angle
with transverse energy $E_T$ then all the kinematic invariants are of order
$E_T^2$. For $E_T (\sim \sqrt{s}/2) \ll M_{Z'}$, the
QCD process dominates. As $E_T$ increases, the $Z'$ contribution
becomes more important, becoming maximal when $E_T \sim M_{Z'}/2$.

Next consider the emission of a soft gluon in the above 
process, i.e. $q(p_1)+\bar q(p_2) \to q'(p_3)+\bar q'(p_4) + g(k)$.
In the soft gluon  and large-$N_c$ approximations, the matrix element is
\begin{eqnarray}
\overline{\sum}\vert{\cal M}\vert^2 &= &
{g_s^6 C_F \over N_c}\; \left( {u^2 + t^2 \over s^2} \right)\;
\Big\{ 2C_F ([13] + [24]) \Big\} \nonumber \\
&& + 2g_s^2 g'^4\; \left({ u^2 + t^2 \over (s-M_{Z'}^2)^2
+ M_{Z'}^2 \Gamma_{Z'}^2} \right)\;
\Big\{ 2C_F ([12] + [34]) \Big\} \; .
\label{example23}
\end{eqnarray}
The distribution of soft gluon radiation is controlled by
the basic antenna pattern (see for example Ref.~\cite{book})
\beq
[ij] \equiv {p_i \cdot p_j \over p_i\cdot k\; p_j \cdot k}
 = { 1 - {\bf n}_i \cdot {\bf n}_j \over
( 1 - {\bf n} \cdot {\bf n}_i)\;
( 1 - {\bf n} \cdot {\bf n}_j)}
\label{antenna}
\eeq
which describes the emission of soft primary gluons with energies $E$:
$\Lambda_{QCD} \ll E \ll E_i, E_j$.
The production of soft hadrons
is then described by extra multiplicative cascading factors
$N'_q(E_T)$ and $  N'_g(E_T)$
within quark and gluon jets respectively (for details see Refs.~\cite{DKMT,drag}): 
\beq
N'_{q,g}(E_T)={ dN_{q,g}\over d \ln E_T }\; ,
\eeq
where $N_{q,g}(E)$ is the particle multiplicity in an individual $q,g$ jet
with energy $E$. Asymptotically 
\beq
N'_{q}(E_T)= {C_F \over N_c} \; N'_{g}(E_T)
\eeq
 with
\beq
{N'_g(E_T)\over N_g(E_T)} = \sqrt{{4 N_c \alpha_s(E_T)\over 2 \pi }}
\; \left[ 1+ O\left( \sqrt{{ \alpha_s\over \pi}}\right) \right]\; .
\label{N_g_formula}
\eeq
Note that in the various asymmetry parameters (e.g.  those used to describe
 the string effect) the cascading factors normally
cancel.
In the present study we will mainly be interested in the
distribution of soft {\it jets} accompanying large-$E_T$
jet production. We shall assume that this is well approximated
by the distribution of a  soft primary gluon which is  given by matrix
elements such as (\ref{example23}).

Returning to the $q \bar q \to q' \bar q'$ process, we see
from (\ref{example23}) that
the radiation pattern is different for the QCD and $Z'$
contributions. At small $E_T$, the distribution is given by the
sum of the $[13]$ and $[24]$ antennas. Soft gluon radiation
is enhanced in the phase space region between the directions
of partons 1 and 3 (and also between 2 and 4). If, however, at large $E_T$
 the $Z'$ contribution is dominant, then the distribution is given
by the sum of the $[12]$ and $[34]$ antennas, which yields a more
symmetric radiation pattern. Thus the soft gluon distribution
acts as a ``partonometer" for probing the hard scattering process.
In general, each type of $2\to 2$ scattering will have its own
distinctive radiation pattern.

In order to make the study more quantitative we must  define
appropriate kinematic distributions and then compute the contributions
of the various subprocesses. In order to fully understand the
differences between these, we first consider the soft gluon
distribution at the parton-parton scattering level with fixed
kinematics. The generic process is
\begin{equation}
a(p_1) + b(p_2)\to c(p_3) + d (p_4) + g(k) \; ,
\label{eq:generic}
\end{equation}
where the  gluon is soft relative to the two 
large-$E_T$ partons $c$ and $d$. Ignoring the gluon momentum in the
energy-momentum constraints, and
using the notation $p^\mu = (E,p_x,p_y,p_z)$, we write
\begin{eqnarray}
p_1^\mu &=& (E_T\cosh\eta, 0, 0, E_T\cosh\eta) \; , \nonumber \\
p_2^\mu &=& (E_T\cosh\eta, 0, 0, -E_T\cosh\eta) \; , \nonumber \\
p_3^\mu &=& (E_T\cosh\eta, 0, E_T, E_T\sinh\eta) \; , \nonumber \\
p_4^\mu &=& (E_T\cosh\eta, 0, -E_T, -E_T\sinh\eta) \; , \nonumber \\
k^\mu &=& (k_T\cosh(\eta+\Delta\eta), k_T\sin\Delta\phi,
 k_T \cos\Delta\phi, k_T\sinh(\eta+\Delta\eta)) \; .
\label{eq:4vecs}
\end{eqnarray}
This is the appropriate form for studying the angular distribution of the soft
gluon jet relative to the large-$E_T$ jet 3, the separation between
these being parametrized by $\Delta\eta$ and $\Delta\phi$. Alternative
variables, more suited to the experimental analysis, are the radial and
polar angle variables in the   ``LEGO plot":
\begin{eqnarray}
\Delta\eta &=&   \Delta R    \cos\beta \; , \nonumber \\
\Delta\phi &=&  \Delta R    \sin\beta \; .
\end{eqnarray}
  which are defined in such a way that
that the LEGO--plot separation between soft jet 
$k$ and hard jet $p_3$ is constant: $R(k,p_3) = \sqrt{\Delta\eta^2+
\Delta\phi^2} = \Delta R$, ($0 \leq \Delta R
< \infty$) , and the
azimuthal orientation of jet $k$ around jet $p_3$ in the LEGO plot
is parametrized by the angle $\beta$, ($0 \leq \beta < 2 \pi$). 

In terms of these variables, the soft gluon phase space is
\begin{equation}
{1\over (2 \pi)^3} \; {d^3 k \over  2 E_k}
 = {1 \over 16\pi^3}\; k_T d k_T \; \Delta R d \Delta R\;  d \beta\; .
\end{equation}
We will be particularly interested in the behaviour of the cross
section as a function of $\beta$ for fixed $k_T, \Delta R$
and fixed $E_T, \eta$.

Note that the $\beta$ distributions presented here should be
considered only as  instructive examples. In the quantitative study
of interference phenomena it may prove useful to exploit other variables,
for example the $\alpha$ variable used in the CDF analysis \cite{exp}.
Another useful discriminator for studying the underlying
scattering dynamics is provided by 
correlations of the type \cite{DKMT,abgn}
\beq
{ N_{UR} \pm N_{LL}- N_{UL} \mp N_{LR}
\over
N_{UR} +N_{LL}+  N_{UL} + N_{LR} } \; ,
\eeq
where $N_{ij}$ is the number of events with the soft jet in the angular region ${ij}$
of the scattering plane ($i$ denotes the upper ($U$) or lower ($L$) quadrant
and $j$ denotes the left ($L$) or  right ($R$) quadrant).

There are a large number of processes of type (\ref{eq:generic}).
The matrix elements are collected in the Appendix. In addition to
the QCD contributions, we include also the contributions 
  from $Z'$ exchange to
$qq\to qq$ scattering (including processes related by crossing),
assuming an interaction of the form
\beq
g'\;  \overline{\psi}_q \gamma_\mu (v_q+a_q\gamma_5) \psi_q\; Z'^{\mu}\; .
\label{zpint}
\eeq
  For illustrative purposes
in our numerical calculations, we use the set of parameters suggested
in Ref.~\cite{Alt} on the basis of a combined fit to LEP and CDF data:
\beq
v_u =1.2\; ,\ \ a_u =3.2\; , \ \ v_d=-1\; ,\ \ a_d = 1\; ,
\eeq
with $g' = e/\sin\theta_W$ and $M_{Z'} = 1$~TeV.

\begin{figure}[tb]
\begin{center}
\mbox{\epsfig{figure=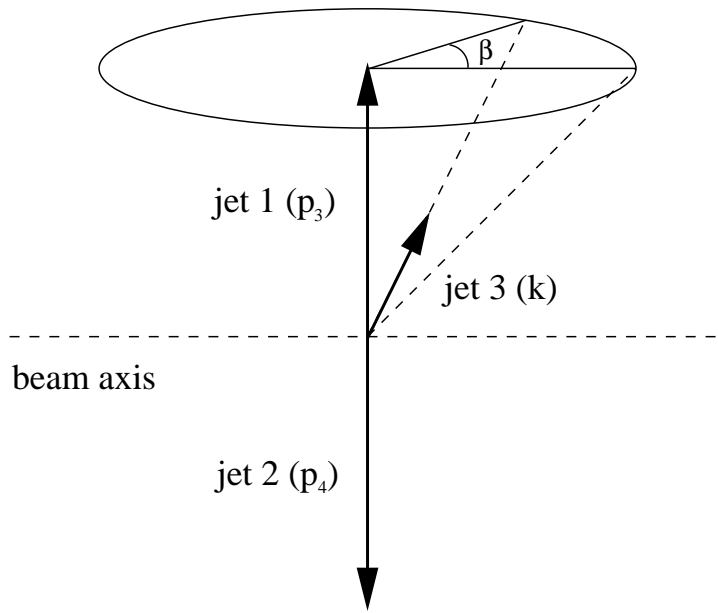,height=8cm}}
\caption{Jet configuration for the numerical studies described in the text.}
\label{fig:lego}
\end{center}
\end{figure}
\begin{figure}[tb]
\begin{center}
\mbox{\epsfig{figure=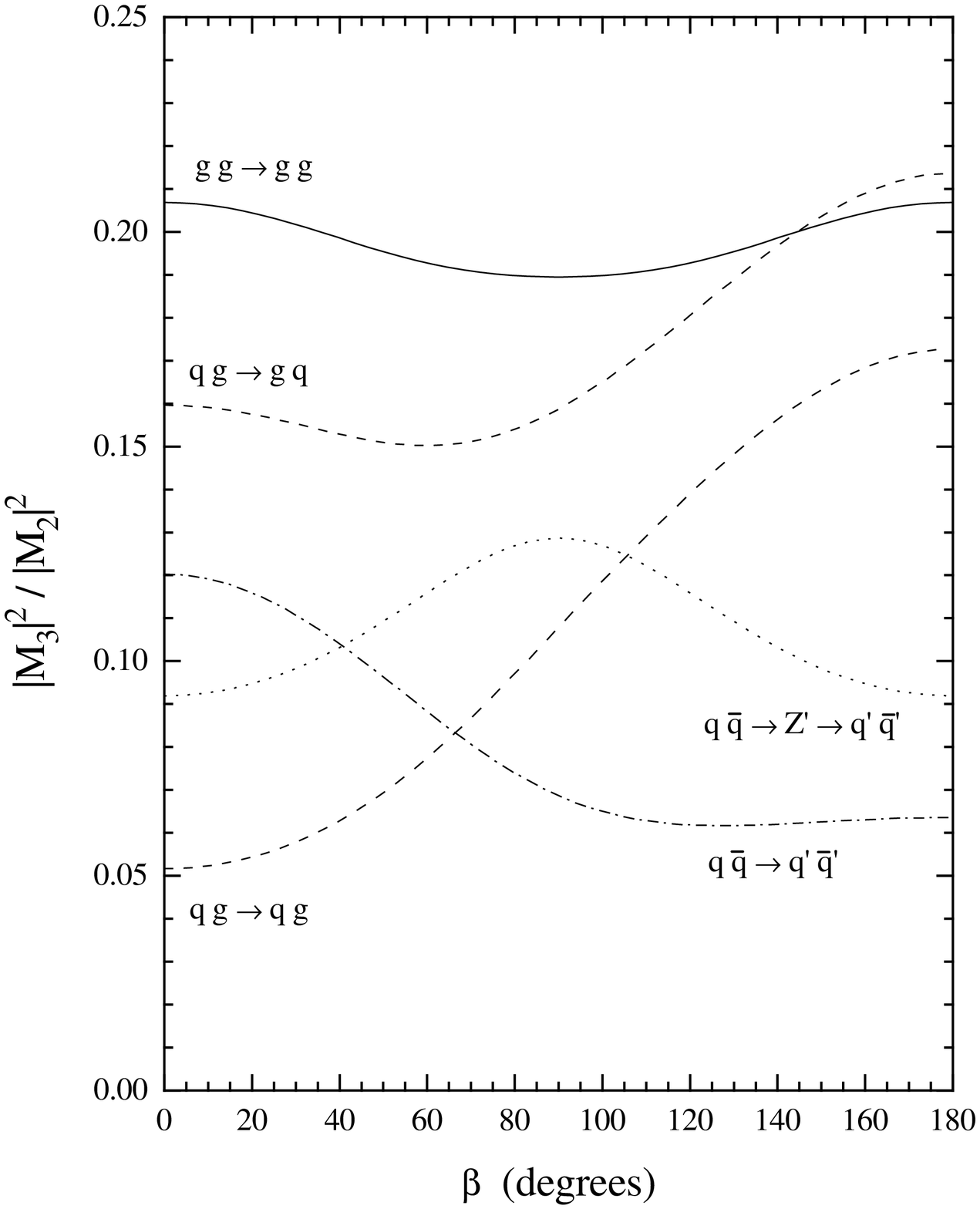,height=12cm}}
\caption{Ratio of the $2\to 3$ and $2 \to 2$ matrix elements as a function
of the soft gluon azimuthal angle about the large-$E_T$ jet, for various
QCD and $Z'$ subprocesses.}
\label{fig:basic2to2}
\end{center}
\end{figure}

We first show results for a very simple kinematic configuration,
$2\to 2$ scattering at $90^{\circ}$ in the parton centre-of-mass
frame, i.e. $\eta=0$ in Eq.~(\ref{eq:4vecs}). We fix $k_T$ and $E_T$
at $10$~GeV and $100$~GeV respectively, so that
$k_T \ll E_T$, and study the angular ($\beta$) distribution of the
soft gluon jet around one of the two large-$E_T$ jets ($p_3$), see
Fig.~\ref{fig:lego}.
Following the \Dzero\ analysis \cite{D0}, we use the variables $\Delta R$ and
$\beta$ defined above. For small $\Delta R$, the radiation pattern 
is dominated by the matrix-element singularity when $k$ and $p_3$ are
parallel: the $\beta$ distribution is constant and simply reflects the
colour charge of the emitting parton.
For large $\Delta R$, the pattern is sensitive to the
colour flow in the scattering  process. We display results for
$\Delta R = 1$, which is typical of the values used in the 
experimental analyses. Figure~\ref{fig:basic2to2} 
shows the soft gluon distribution
for various $2\to 2$ scattering processes as 
a function of $\beta$\footnote{For the choice of kinematics adopted here,
the distribution is symmetric in $ \beta \leftrightarrow  2\pi - \beta$.}.
The quantity plotted is the ratio of the $2 \to 3$ and $2\to 2$ 
matrix elements, omitting an overall factor of  $g_s^2$. 
The characteristics of these distributions can be understood in terms
of the colour strings linking the initial- and final-state partons,
see for example Ref.~\cite{book}. Notice in particular that the 
distribution for the $gg\to gg$ process is approximately constant, 
reflecting the comparable contributions to the radiation pattern
from the incoming and outgoing gluons. For the $qg \to gq $
and $qg \to qg$ processes, the distributions are peaked in the
backward ($\cos\beta <0$) direction. For the former, this corresponds
to the colour string between an outgoing $8$ and $\overline{8}$, 
as for $gg\to gg$. 

\begin{figure}[tb]
\begin{center}
\mbox{\epsfig{figure=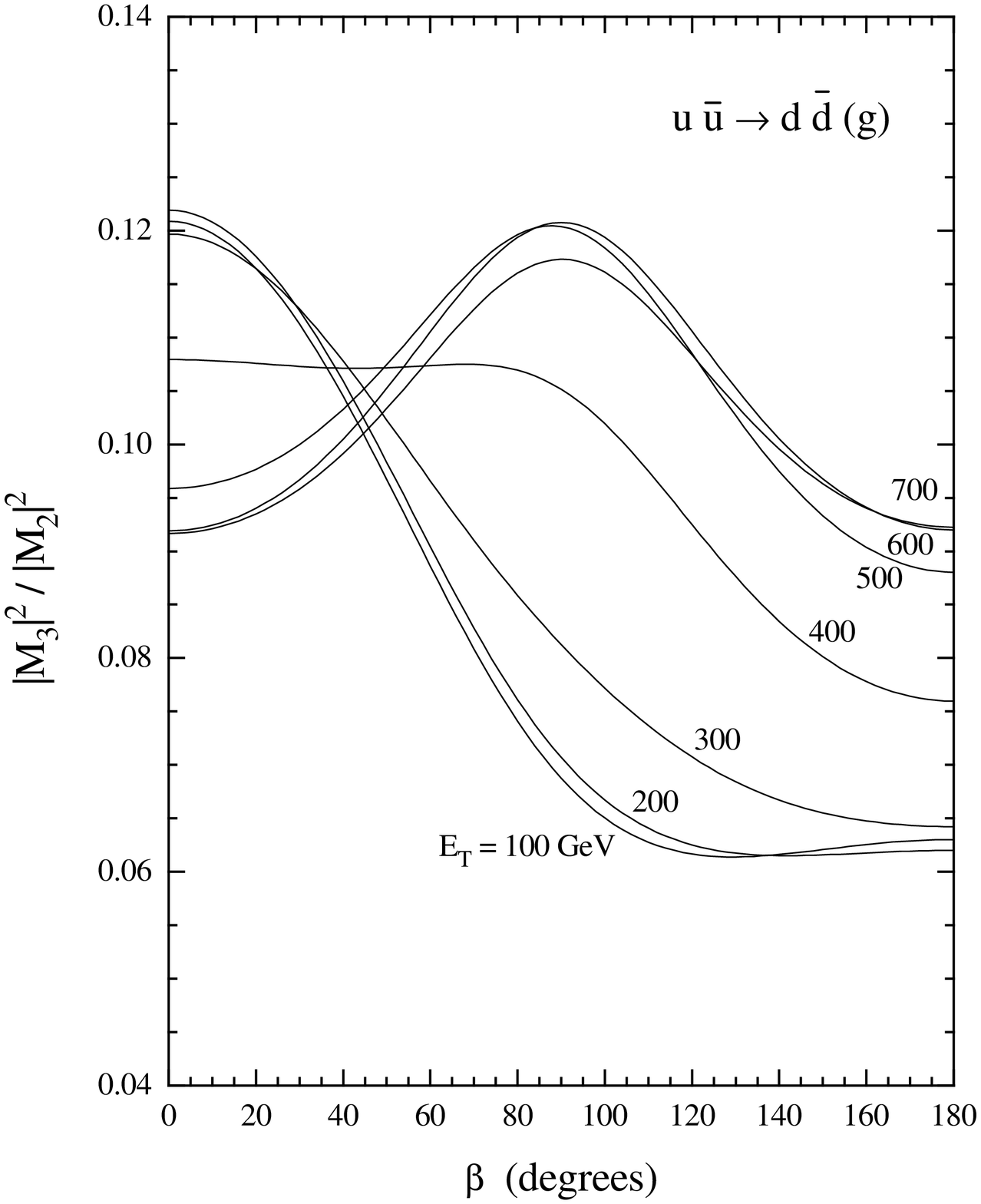,height=12cm}}
\caption{Ratio of the $2\to 3$ and $2 \to 2$ matrix elements as a function
of the soft gluon azimuthal angle about the large-$E_T$ jet for the 
$u \bar u \to d \bar d (g)$ subprocess, including QCD and $Z'$ contributions.}
\label{fig:qcdzp}
\end{center}
\end{figure}

Also shown in Fig.~\ref{fig:basic2to2} is the radiation pattern
for the process $ q \bar q \to Z' \to q' \bar{q}'$. In this case
the distribution has the simple analytic form
\begin{equation}
\frac{d\cW}{d\beta} = 2C_F( [12]+[34] ) = \frac{4C_F}{k_T^2}
\;\left[ 1 +  { 1 \over \cosh^2(\Delta R\cos\beta)
- \cos^2(\Delta R\sin\beta)       }\right]\; .
\label{pat1234}
\end{equation}
Note that the $[12]$ antenna corresponding to gluon radiation
from the initial-state $ q\bar q$ gives a contribution which 
is {\it independent} of $\beta$. 
In the large-$N_c$ limit, the corresponding QCD process gives
the distribution
\begin{equation}
\frac{d\cW}{d\beta} \to 2C_F( [13]+[24] ) = \frac{4C_F}{k_T^2}
\;\left[  { \cosh^2(\Delta R\cos\beta) + 
 \sinh(\Delta R\cos\beta)\cos(\Delta R\sin\beta)
  \over \cosh^2(\Delta R\cos\beta)
- \cos^2(\Delta R\sin\beta)       }\right]\; .
\end{equation}
In contrast to (\ref{pat1234}), this distribution is not symmetric
about  $\beta = 90^{\circ}$.
For a heavy $Z'$ with 
sizeable couplings to quarks, we would expect
a transition from the ``QCD" radiation pattern at low $E_T$ to the 
``$Z'$" radiation pattern at high $E_T$. This is illustrated in
Fig.~\ref{fig:qcdzp},
for the parton scattering process $u \bar u \to d \bar d$. The 
distribution is obtained from Eq.~(\ref{qqbqqbp}) and 
contains the contributions from the QCD and $Z'$ processes,
and their interference. For these simple kinematics, $E_T
= M_{Z'}/2 = 500$~GeV
corresponds to the resonance peak where the $Z'$ contribution is maximal.
Notice the significant  change in the shape of the distribution 
as $E_T$ approaches this value from below. 

\begin{figure}[tb]
\begin{center}
\mbox{\epsfig{figure=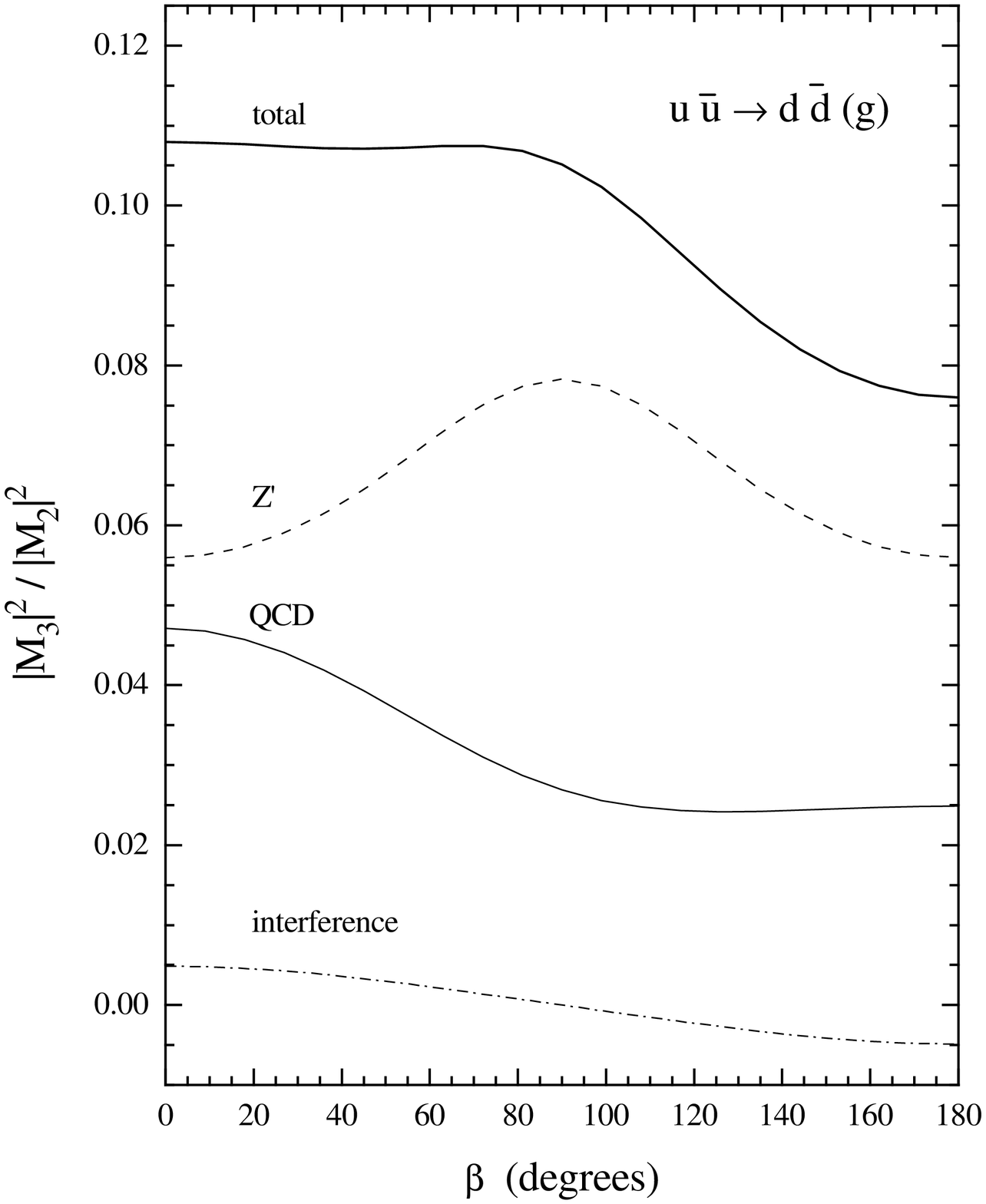,height=12cm}}
\caption{Decomposition of  the 
$u \bar u \to d \bar d (g)$ radiation pattern of Fig.~\protect\ref{fig:qcdzp},
for $E_T = 400$~GeV. The curves represent the various contributions to the
$2\to 3$ matrix element squared, normalized in each case to the total QCD$+Z'$
$2\to 2$ matrix element squared. }
\label{fig:qcdzpint}
\end{center}
\end{figure}

The interference between the QCD and $Z'$ contributions
 deserves further discussion. By colour conservation, this is absent
in the leading-order (no gluon radiation) amplitudes, and at 
$O(\alpha_s)$ is suppressed by  a single power of $N_c$ in the 
large-$N_c$ limit, see Eq.~(\ref{qqbqqbp}). The corresponding antenna
pattern is given by 
\beq
\frac{d\cW_{\rm int}}{d\beta} \to 2 C_F ([13] + [24] - [14] - [23]) 
 = \frac{8C_F}{k_T^2}
\;\left[  { 
 \sinh(\Delta R\cos\beta)\cos(\Delta R\sin\beta)
  \over \cosh^2(\Delta R\cos\beta)
- \cos^2(\Delta R\sin\beta)       }\right]\; ,
\label{intpat}
\eeq
where we have substituted the kinematic variables relevant to 
Fig.~\ref{fig:qcdzp}, i.e. $\eta = 0$. The interference contribution vanishes on
the $Z'$ resonance (i.e. for $E_T = 500$~GeV in Fig.~\ref{fig:qcdzp})
and is everywhere numerically small relative to
the sum of the QCD and $Z'$ amplitudes squared. As an illustration,
Fig.~\ref{fig:qcdzpint} shows the decomposition of the $u \bar u \to d \bar
d(g)$ radiation pattern for $E_T = 400$~GeV. For this kinematic
configuration ($90^\circ$ scattering in the parton centre-of-mass frame)
the interference is antisymmetric about $\beta = 90^\circ$.
Note  that the antenna pattern in Eq.~(\ref{intpat}) is asymmetric with respect
to the interchange $ p_3 \leftrightarrow  p_4$ and therefore  vanishes if one 
does not distinguish
the final-state jets. It also vanishes after integration
over the angles between the $q\bar q$ and $q'\bar{q}'$ antennae,
see Ref.~\cite{KSt}. This can readily be seen by writing Eq.~(\ref{intpat})
as
\beq
 2 C_F ([13] + [24] - [14] - [23]) 
 =  - \frac{8C_F}{\vert \vec{k}\vert^2}
\;\left[  {  \cos\phi 
  \over \sin\theta_1 \; \sin\theta_3    }\right]\; ,
\label{intpatbis}
\eeq
where  $\theta_{1,3}$ is the polar angle between the $\vec{k}$ and 
 and $\vec{p}_{1,3}$ vectors,
and $\phi$  is the azimuthal angle between the planes formed
by $\vec{k}, \vec{p}_1$ and  $\vec{k}, \vec{p}_3$. 

More generally, the interference
between diagrams with different colour flows at the amplitude
level destroys the factorization of the radiation pattern into
a product of  the eikonal antenna factors and the lowest-order cross sections.
Note that an  analogous interference contribution induces  
``colour interconnection effects" in the pattern of gluon radiation
accompanying $e^+e^- \to  q_1 \bar{q}_2 q_3 \bar{q}_4$ events, see 
Ref.~\cite{SKMW}. 
%Despite the fact that this effect is colour suppressed, i.e. 
%$O(N_c^{-2})$, it affects the precise determination of the $W$ mass
%from such events.

\begin{figure}[tb]
\begin{center}
\mbox{\epsfig{figure=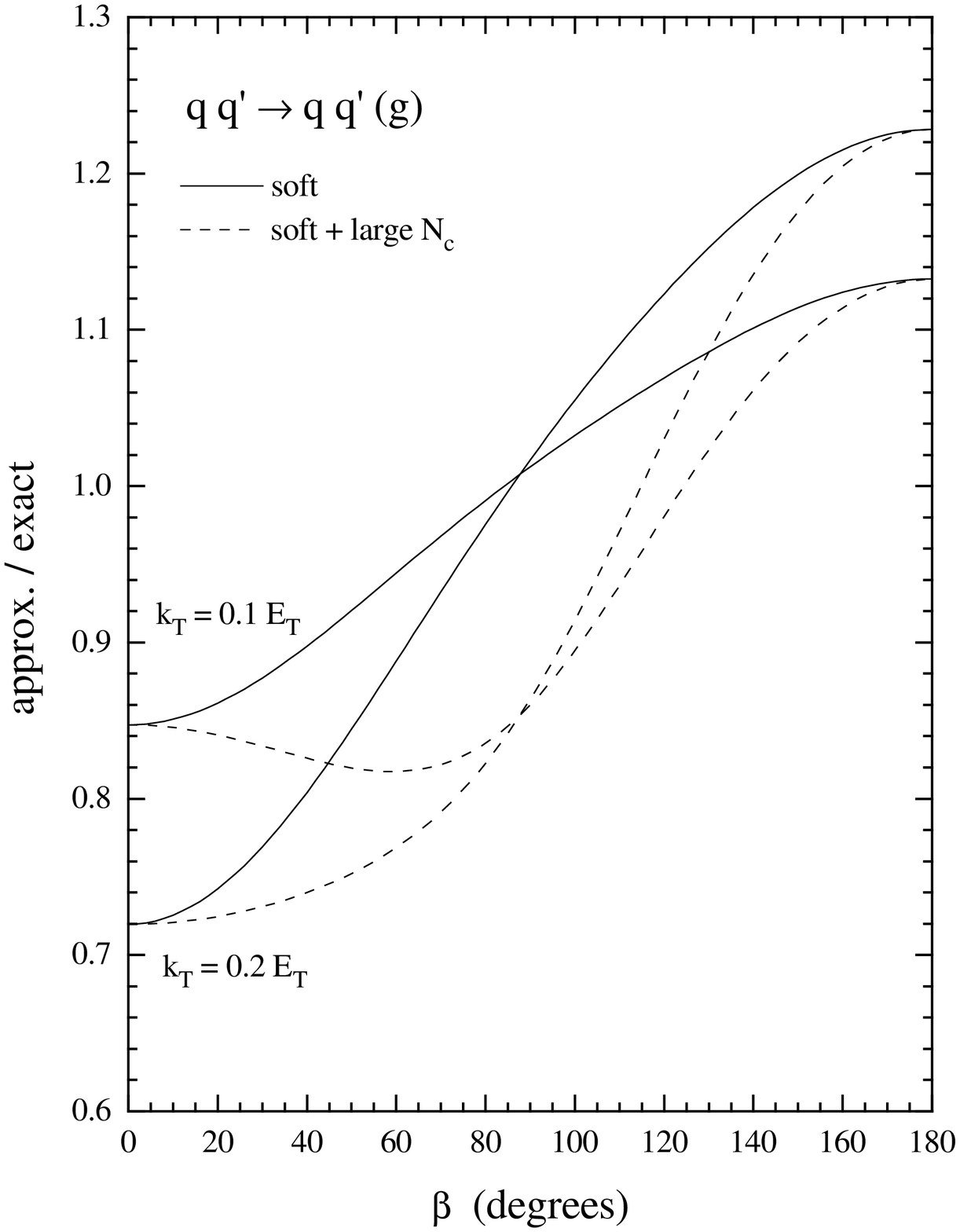,height=12cm}}
\caption{Ratio of approximate and exact matrix elements, as described
in the text.}
\label{fig:approx}
\end{center}
\end{figure}

We conclude this section with some remarks
on the validity of the soft gluon and large-$N_c$ approximations.
Consider, for example, the QCD $2\to 3$ process $q q' \to q q' g$.
The exact matrix element is \cite{exact2to3}
\beq
\overline{\sum}\vert{\cal M}\vert^2(\mbox{exact}) = 
{g_s^6 C_F \over N_c}\; \left( {s^2 + s'^2 +u^2 + u'^2\over 2 t t'} \right)\;
\Big\{ 2C_F ([14] + [23]) + \frac{1}{N_c} [12;34]\Big\} \; ,
\label{eq:exact}
\eeq
where
\bea
s = (p_1+p_2)^2, & t = (p_1-p_3)^2, & u = (p_1-p_4)^2, \nonumber \\
s' = (p_3+p_4)^2, & t' = (p_2-p_4)^2, & u' = (p_2-p_3)^2.
\eea
The soft gluon approximation which we have used in the above calculations
corresponds to 
\beq
\overline{\sum}\vert{\cal M}\vert^2(\mbox{soft}) = 
{g_s^6 C_F \over N_c}\; \left( {s^2 +u^2 \over  t^2 } \right)\;
\Big\{ 2C_F ([14] + [23]) + \frac{1}{N_c} [12;34]\Big\} \; ,
\label{eq:sapprox}
\eeq
with $s = s'$ etc.   Making the large-$N_c$ approximation, 
the above expression simplifies further to
\beq
\overline{\sum}\vert{\cal M}\vert^2(\mbox{soft, large-}N_c) = 
{g_s^6 C_F \over N_c}\; \left( {s^2 +u^2 \over  t^2 } \right)\;
\Big\{ 2C_F ([14] + [23]) \Big\} \; .
\label{eq:sNapprox}
\eeq
To test the validity of these approximations, we consider the ratios
of the matrix elements in (\ref{eq:sapprox}) and (\ref{eq:sNapprox}) to that 
in (\ref{eq:exact}).
For the latter, we use exact $2\to 3$ kinematics defined by 
\begin{eqnarray}
p_1^\mu &=& (E, 0, 0, E) \; , \nonumber \\
p_2^\mu &=& (E, 0, 0, -E) \; , \nonumber \\
p_3^\mu &=& (E_T\cosh\eta, 0, E_T, E_T\sinh\eta) \; , \nonumber \\
p_4^\mu &=& (E_4, -k_T\sin\Delta\phi, -E_T
- k_T \cos\Delta\phi, -E_T\sinh\eta - k_T\sinh(\eta+\Delta\eta) \; , \nonumber \\
k^\mu &=& (k_T\cosh(\eta+\Delta\eta), k_T\sin\Delta\phi,
 k_T \cos\Delta\phi, k_T\sinh(\eta+\Delta\eta)) \; .
\label{eq:4vecsexact}
\end{eqnarray}
with $E_4 = \vert \mbox{\bf p}_4\vert$ and $2E=E_3 + E_4 + E_k$.
Figure~\ref{fig:approx} shows the ratio of the approximate
and exact matrix elements as a function of $\beta$ for $\eta=0$,
$\Delta R = 1$, and  $k_T/E_T= 0.1, 0.2$. 
The solid lines  indicate that the  corrections to the radiation pattern in
the soft approximation scale as $O(k_T/E_T)$. The difference between
the dashed  and solid lines is $O(10\%)$, consistent with 
$O(1/N_c^2)$ corrections to the large-$N_c$ approximation.
The  curves in Fig.~\ref{fig:approx} attain their maximum deviation
from unity at the end-points, $\cos\beta = \pm 1$, since
 for this choice of kinematics the ratio of the gluon
 energy to that of jet 3 is $E_k/E_3 = \cosh({\cos\beta})k_T/E_T$,
which for fixed $k_T / E_T$ is maximal at large $\vert\cos\beta\vert$.
In what follows we shall use the soft gluon matrix elements
listed in the Appendix, retaining the complete set of antennas valid
for arbitrary $N_c$.

\section{Convolution with Parton Distributions}
\label{sec:sec3}

\begin{figure}[tb]
\begin{center}
\mbox{\epsfig{figure=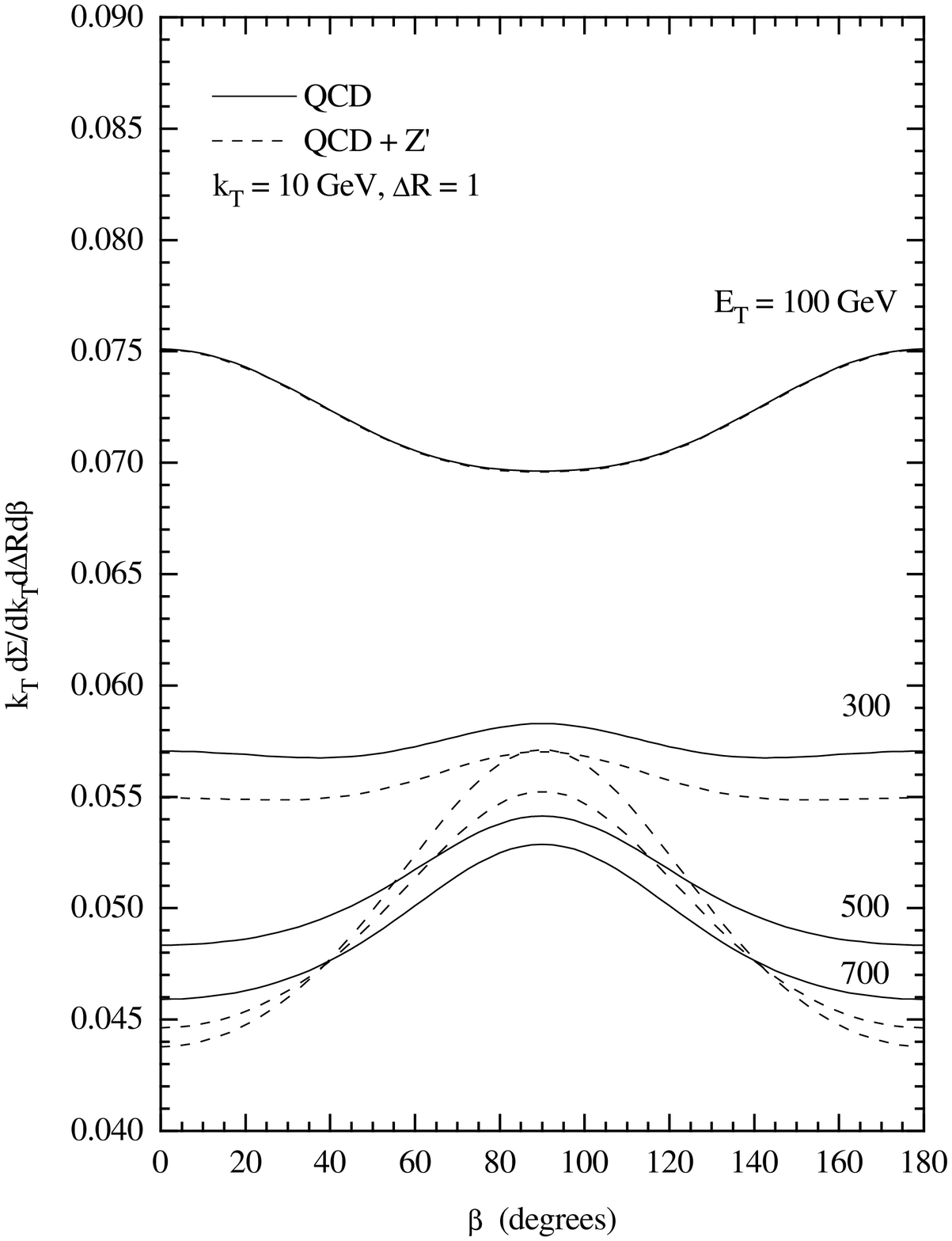,height=12cm}}
\caption{Radiation pattern for the soft gluon jet in
large-$E_T$ dijet production in $\bar p  p$ collisions
at 1.8~TeV. All scattering processes are included, weighted
by the appropriate parton distributions.}
\label{bigone}
\end{center}
\end{figure}

To obtain realistic predictions for the distribution of soft
gluon radiation accompanying large-$E_T$ jets at, say, the 
Tevatron $\bar p p$ collider, we must convolute the matrix elements
with appropriate parton distributions. This then gives a $E_T$-dependent
mixture of the various distributions shown in Fig.~\ref{fig:basic2to2}.
To avoid unnecessary complications from kinematics, and to make
contact with the parton-level results obtained above, we
consider the production of a pair of large-$E_T$ jets at zero
rapidity in the laboratory frame, i.e. $E_{T3} = E_{T4} = E_T$,
$\eta_3 = \eta_4 = 0$. The leading-order inclusive two-jet 
cross section is 
\bea
{ d^3\sigma \over d\eta_3 d \eta_4 d E_T } &=& {E_T \over 8\pi x_1x_2 s^2}
\; \sum_{a,b,c,d=q, \bar q, g} f_a(x_1,\mu^2) f_b(x_2,\mu^2) \nonumber \\
& & \times\; {1\over 1 + \delta_{cd} } \; 
 \overline{\sum}\vert{\cal M}(ab\to cd)\vert^2\; ,
\label{eq:sig2}
\eea
with $x_1 = x_2 = x_T \equiv 2 E_T/\sqrt{s}$ for our choice 
of kinematics. The symmetry factor $\delta_{cd}$ is $1(0)$ for identical
(non-identical) partons in the final state. Next-to-leading order
corrections to the inclusive cross section 
are approximately taken into account by the scale choice
$\mu = E_T/2$. We use the recent MRS(R2) parton set ($\alpha_s(M_Z^2)
= 0.120$), which gives a  good overall description of the 
CDF and \Dzero\ large-$E_T$ inclusive 
jet cross section~\cite{MRSR}. The corresponding two-jet $+$ soft gluon
inclusive cross section is
\bea
{ d^6\sigma \over d k_T d \Delta R d \beta  d\eta_3 d \eta_4 d E_T } 
&=& {k_T \Delta R E_T \over 128\pi^4 x_1x_2 s^2}
\; \sum_{a,b,c,d=q, \bar q, g} f_a(x_1,\mu^2) f_b(x_2,\mu^2) \nonumber \\
& & \times\; {1\over 1 + \delta_{cd} } \; 
 \overline{\sum}\vert{\cal M}(ab\to cd+g)\vert^2\; .
\label{eq:sig3}
\eea
For the additional factor of $\alpha_s$ coming from the soft gluon emission
we use $\mu = k_T$. 
As before, we consider the ratio of the cross sections in (\ref{eq:sig3})
and (\ref{eq:sig2}) as a function of the soft gluon variables:
\beq
{d\Sigma \over d k_T d \Delta R d \beta} \equiv
\left[ { d^3\sigma \over d\eta_3 d \eta_4 d E_T }\right]^{-1}
\; { d^6\sigma \over d k_T d \Delta R d \beta  d\eta_3 d \eta_4 d E_T }\; .
\eeq
Figure~\ref{bigone} shows the distribution $k_T d\Sigma /
 d k_T d \Delta R d \beta
 $ as a function of $\beta$ for
 $k_T = 10$~GeV, $\Delta R = 1$ and various values of $E_T$.
 The solid curves are the QCD predictions, and the dashed curves
 include also the  $Z'$ contribution with same parameters as
 in the previous figures. We note the following.
 
\begin{figure}[tb]
\begin{center}
\mbox{\epsfig{figure=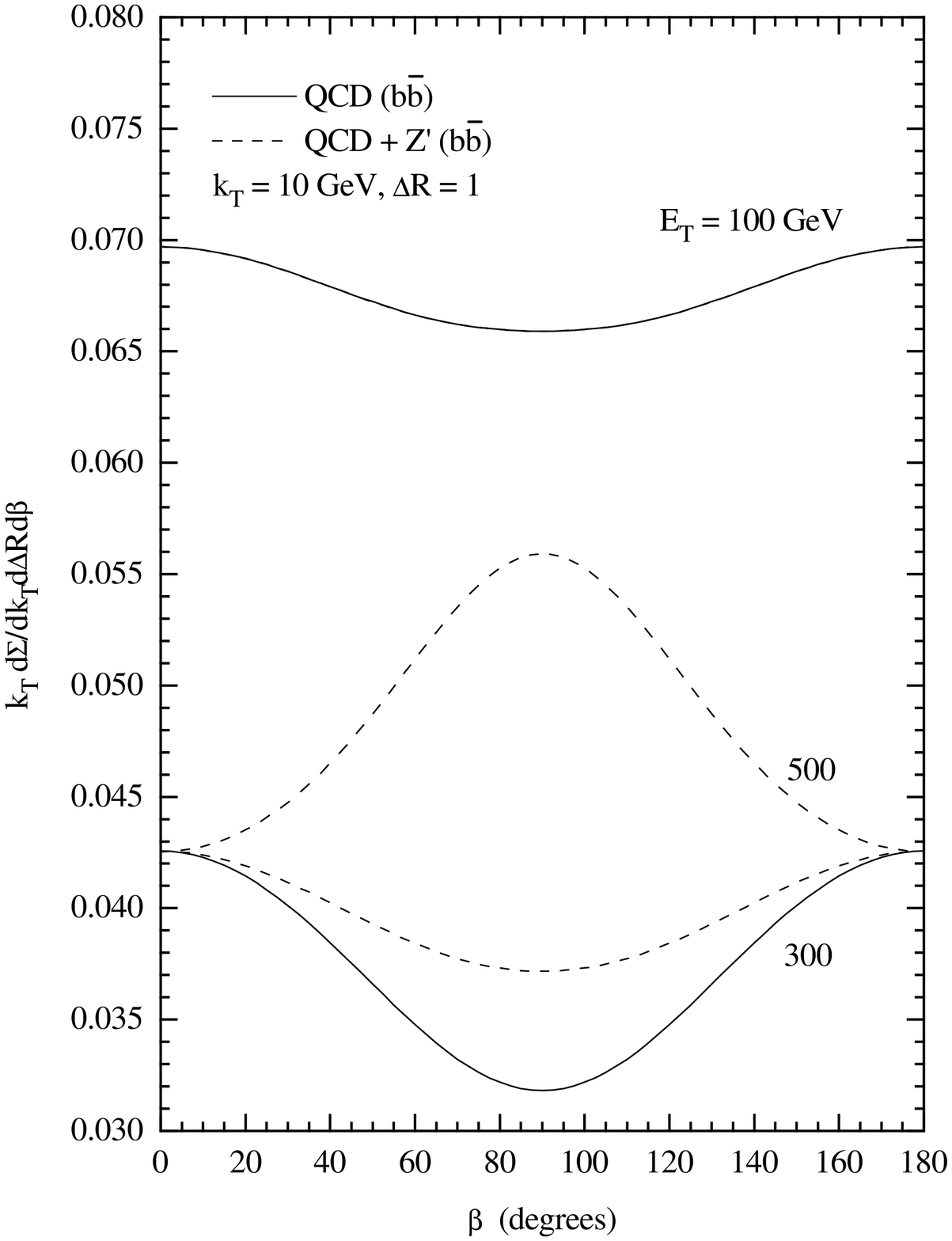,height=12cm}}
\caption{As for Fig.~\protect\ref{bigone}, except that only the large-$E_T$
$b \bar{b}$ final state is included. The QCD (solid) 
curves for $E_T = 300$~GeV and $500$~GeV are indistinguishable.}
\label{bigoneb}
\end{center}
\end{figure}

\begin{itemize}
\item[{(i)}] All the distributions are symmetric about $\beta  = 90^\circ$.
This is due to
our choice of symmetric  large-$E_T$ jet kinematics, and
 because we have summed over all combinations of  partons (jets)
in the initial and final states, e.g. $qg, gq \to qg, gq$.\footnote{By varying
the jet rapidities one can preferentially select, say, the $qg$ initial
state over the $gq$ initial state and expose the asymmetries of the
$qg$ radiation pattern in Fig.~\protect\ref{fig:basic2to2}.}
\item[{(ii)}]
For the QCD curves, the radiation pattern decreases in magnitude as $E_T$
increases. This is simply a colour charge effect: as one moves from small $E_T$ to
large $E_T$ the dominate subprocess scattering changes from $gg$ to
$qg$ to $q \bar q$.
\item[{(iii)}] The effect of the $Z'$ contribution is clearly visible
in the dashed curves at large $E_T$. As already 
seen in Fig.~\ref{fig:basic2to2}, the $Z'$ radiation pattern   
has a peak at $\beta = 90^\circ$. The effect is most
significant for $E_T = M_{Z'}/2$, as expected.
 But note also a similar (though weaker)
peak structure in the QCD `background'. This arises from the 
$q \bar q \to q \bar q +g$ process which, for the $t$-channel
gluon exchange contribution,  has a dominant $C_F([12]+[34])$
antenna pattern as for $q \bar q \to Z' \to q \bar q$. The effect
only arises for the identical quark process -- there is no such
$t$-channel contribution for $q \bar q \to q' \bar{q}'+g$.
\item[{
(iv)}] This latter comment suggests a method for further
enhancing the $Z'$ signal, or other new physics contributions,
 over the QCD background. Figure~\ref{bigoneb}
shows the effect of selecting only the $ b\bar{b}$ component of
the final state in Fig.~\ref{bigone}. Such a selection could
in principle be achieved experimentally by means of a vertex detector.
At small $E_T$ the dominant process is $gg\to b \bar{b}$ with a large,
approximately flat $\beta$ distribution. At large $E_T$, the QCD 
process $q \bar q\to b \bar{b}$ has a dominant antenna pattern given by
 $C_F([13]+[24])$, which is very different from the $C_F([12]+[34])$
pattern of $q \bar q\to Z' \to  b \bar{b}$. We note in passing that
purely in the context of QCD, a $b \bar{b}$ subsample of the 
large-$E_T$ dijet events should exhibit a markedly different distribution of
soft hadronic radiation.
\end{itemize}

\section{Application to Composite Models}
\label{sec:sec4}

If the quarks and leptons of the Standard Model are composite
objects, one would expect to see new four-fermion contact interactions
well below the scale $\Lambda$ characterizing the size of the composite
states \cite{ELP}. In particular, a four-quark contact interaction would
produce a flattening of the large-$E_T$ jet distribution at scales
$E_T \sim \Lambda$. The colour structure of such a contact interaction
would generate a distinctive antenna pattern of soft radiation, which could
be used to distinguish between standard QCD and different types of contact
interactions between quarks.

It is straightforward to write down the most general 
SU(3)$\times$SU(2)$\times$U(1) invariant four-quark contact interaction,
see for example Ref.~\cite{EHLQ}.
The generic form is
\beq
{\cal L}_c = {4\pi\over 2\Lambda^2}\; \sum\; \bar q \gamma^\mu q \; 
\bar q \gamma_\mu q \; ,
\label{generic}
\eeq
where the sum is over the colour, flavour and helicity labels carried
by the quark fields.  There are two possible colour structures for
such interactions:
\beq
\bar{q}_i \gamma^\mu q_i\;  \bar{q}_j \gamma_\mu q_j, \ \ 
\bar{q}_i T^a_{ik} \gamma^\mu q_k\;  \bar{q}_j T^a_{jl} \gamma^\mu q_l\; ,
\label{colourstruct}
\eeq
where $i,j,k,l=1,2,3$ are colour labels and the $T^a_{ij}$ ($a=1,..,8$)
are the  SU(3) colour matrices of QCD.
The two terms in Eq.~(\ref{colourstruct}) correspond to colour singlet
and octet exchange respectively.
Combining these new contact interactions with those of standard QCD,
one obtains additional contributions to the $2\to 2$ scattering
processes involving quarks. For example, for the  $qq\to qq$
amplitude squared,  the additional contributions are or order
$\Lambda^{-4}$ (contact interaction squared) and $(t \Lambda^2)^{-1}$, 
$(u \Lambda^2)^{-1}$ for the QCD-contact interference.

The additional contributions to the radiation pattern can be calculated
in the same way. As a specific example, we consider the most widely
used interaction of type (\ref{generic}), involving the 
product of left-handed, colour-singlet, isoscalar first-generation quark
currents:
\beq
{\cal L}_c = {4\pi\eta\over 2\Lambda^2}\; \sum_{i,j=1,3}\; 
\bar{q}_{iL} \gamma^\mu q_{iL} \; \bar{q}_{jL} \gamma_\mu q_{jL}\; , 
\eeq
where $q=(u,d)$ and $\eta = \pm 1$ determines whether the 
interference with the standard QCD interaction is constructive or
destructive. In fact the effect of such an interaction
can be readily obtained as a limiting case of the
 $Z'$ results listed in the Appendix, specifically by setting
 $v' =-a' = 1$ and taking the limit
\beq
M_{Z'}, g' \to \infty\quad \mbox{with} \quad 
\left({2 g' \over  M_{Z'}}\right)^2 = -{4 \pi  \eta \over \Lambda^2} \quad 
\mbox{fixed}.  
\eeq
Thus for $ud \to ud+g $ we obtain (cf.~\ref{qqpqqp})
\begin{eqnarray}
\overline{\sum}\vert{\cal M}\vert^2 &= &
{g_s^6 C_F \over N_c}\; \left( {s^2 + u^2 \over t^2} \right)\;
\Big\{ 2C_F ([14] + [23]) + \frac{1}{N_c} [12;34]\Big\} \nonumber \\
&& + \;  {(4\pi)^2 g_s^2 s^2\over \Lambda^4}
\; \Big\{ 2C_F ([13] + [24]) \Big\}\; + \;  
{ 4 \pi \eta g_s^4 s^2 \over  t\Lambda^2} 
  \frac{C_F}{N_c} 
\; \Big\{ 2 ([14] + [23] - [12]-[34]) \Big\} \; .  \nonumber \\ 
\end{eqnarray}
Radiation patterns for the other quark interactions can be obtained
in a similar way. The numerical results will be qualitatively similar to 
those for the heavy  $Z'$ presented in Sections \ref{sec:sec2} and 
\ref{sec:sec3}. 

For a four-quark interaction corresponding to {\it colour octet}
exchange, the effect on the radiation pattern at large $E_T$ will
be much less dramatic, since the antenna structure for, say,
$q \bar q\to q \bar q$ scattering will be the same for both
standard gluon exchange and for the contact interaction. In principle,
therefore, the radiation pattern provides a unique method for
unravelling the colour structure of a new four-quark interaction.

\section{Conclusions}

We have demonstrated in this paper how the hadronic antenna 
pattern due to gluon radiation may be used as a diagnostic tool
which reveals the short-distance colour flow dynamics, and hence may be
used to discriminate between different production mechanisms for
large-$E_T$ jets in ${\bar p} p$ or $pp$ collisions. In particular, we
have discussed in some detail the hadronic antenna pattern
associated with the production of a $Z'$ boson decaying into
jet pairs, and contrasted it with that predicted by 
conventional QCD. We have also related
contact interactions with different colour exchanges to
the QCD and $Z'$ cases, so that they may also be distinguished
by their hadronic antenna patterns.

The diagnostic tool provided by these hadronic antenna 
patterns may be useful in the analysis of current large-$E_T$
jet events at the Fermilab ${\bar p} p$. If they turn out to
have some component beyond conventional QCD, gluon radiation
may help pin down the additional production mechanism. Even
if the current large-$E_T$ data turn out to be satisfactorily
fitted within conventional QCD, this tool may be a useful
addition to the kit of analysis methods to be applied to
future hadronic-jet data at even larger $E_T$ at the LHC.

Finally, we note that the technique of using hadronic antenna patterns 
as a probe  of new physics has many other applications other than 
those discussed in detail in this study. For example, in
Higgs production by gluon-gluon fusion
 at hadron colliders the density of hadrons
 in the plateau corresponding to the incoming partons (the $[12]$ antenna)
 should be approximately twice as large as for $Z'$ production.
Another application is to final states in deep-inelastic scattering, where
the technique could be used, for example, 
 as a probe of ``rapidity-gap" physics.

\vspace{0.5cm}
\noindent {\bf Acknowledgements} \\
We thank G.~Blazey, Yu.~Dokshitzer, M.~Mangano and N.~Varelas
for useful comments and discussions. The work of J.E. was
supported in part by the Director, Office of Energy Research,
Office of Basic Energy Science of the U.S. Department of
Energy, under Contract DE-AC03-76SF00098. The work of VAK and WJS was
supported in part by the UK PPARC and the EU Programme
``Human Capital and Mobility'', Network ``Physics at High Energy
Colliders'', contract CHRX-CT93-0357 (DG 12 COMA).

\vspace{0.5cm}

\section*{Appendix}
\setcounter{equation}{0}
\setcounter{section}{1}
\renewcommand{\theequation}{\Alph{section}\arabic{equation}}

The following expressions are for the spin/colour summed/averaged 
matrix elements for the scattering processes
\begin{equation}
a(p_1) + b(p_2)\to c(p_3) + d (p_4) + g(k)
\end{equation}
in the soft gluon approximation. The QCD matrix elements
are taken from Refs.~\cite{DKT,DKTSJNP,emw}. The antennae are defined by
\begin{eqnarray}
[ij] & = & {p_i \cdot p_j \over p_i \cdot k\; p_j \cdot k}\; ,\nonumber\\ [0pt]
[ij;kl] & = & 2[ij] + 2[kl] -[ik]-[il]-[jk]-[jl] \; .
\end{eqnarray}
and
\begin{equation}
s=(p_1+p_2)^2\; ,\quad t=(p_1-p_3)^2\; ,\quad u=(p_1-p_4)^2\; .
\end{equation}
The $Z'$ couplings are defined in Eq.~(\ref{zpint}), with $(v,a)$ and
$(v',a')$ denoting the vector, axial couplings of the $Z'$
to quarks $q$ and $q'$ respectively.  Only a subset of 
all the possible processes are listed; the rest can be obtained by
crossing.

\subsubsection*{$q(p_1)+\bar q(p_2) \to q'(p_3)+\bar q'(p_4) + g(k)$}
\begin{eqnarray}
\overline{\sum}\vert{\cal M}\vert^2 &= &
{g_s^6 C_F \over N_c}\; \left( {u^2 + t^2 \over s^2} \right)\;
\Big\{ 2C_F ([13] + [24]) + \frac{1}{N_c} [14;23]\Big\} \nonumber \\
&& + 2g_s^2 g'^4\; \left({ (u^2 + t^2)\{(v^2+a^2)(v'^2+a'^2) \}
+(u^2-t^2)\{4 vav'a'  \} \over (s-M_{Z'}^2)^2
+ M_{Z'}^2 \Gamma_{Z'}^2} \right) \nonumber \\
&&\times
\; \Big\{ 2C_F ([12] + [34]) \Big\} \nonumber \\
&& + \frac{g_s^4 g'^2}{N_c}\; 2 \Re \left({ (u^2 + t^2)\{vv' \}
+(u^2-t^2)\{aa'  \} \over s [  s-M_{Z'}^2
+ i M_{Z'} \Gamma_{Z'} ] } \right) \nonumber \\
&& \times
\; \Big\{ 2 C_F ([13] + [24] - [14] - [23]) \Big\}  \nonumber \\
\label{qqbqqbp}
\end{eqnarray}

\subsubsection*{$q(p_1)+q'(p_2) \to q(p_3)+q'(p_4) + g(k)$}

\begin{eqnarray}
\overline{\sum}\vert{\cal M}\vert^2 &= &
{g_s^6 C_F \over N_c}\; \left( {s^2 + u^2 \over t^2} \right)\;
\Big\{ 2C_F ([14] + [23]) + \frac{1}{N_c} [12;34]\Big\} \nonumber \\
&& + 2g_s^2 g'^4\; \left({ (s^2 + u^2)\{(v^2+a^2)(v'^2+a'^2) \}
+(s^2-u^2)\{4 vav'a'  \} \over (t-M_{Z'}^2)^2
+ M_{Z'}^2 \Gamma_{Z'}^2} \right) \nonumber \\
&&\times
\; \Big\{ 2C_F ([13] + [24]) \Big\} \nonumber \\
&& + \frac{g_s^4 g'^2}{N_c}\; 2\Re \left({ (s^2 + u^2)\{vv' \}
+(s^2-u^2)\{aa'  \} \over t[ t-M_{Z'}^2
+  i M_{Z'} \Gamma_{Z'} ] } \right) \nonumber \\
&& \times
\; \Big\{ 2C_F ([14] + [23] - [12]-[34]) \Big\}    \nonumber \\
\label{qqpqqp}
\end{eqnarray}

\subsubsection*{$q(p_1)+q(p_2) \to q(p_3)+q(p_4) + g(k)$}

\begin{eqnarray}
\overline{\sum}\vert{\cal M}\vert^2 &= &
{g_s^6 C_F \over N_c}\; \left( {s^2 + u^2 \over t^2}
-\frac{1}{N_c}\; {s^2\over  tu} \right)\;
\Big\{ 2C_F ([14] + [23]) + \frac{1}{N_c} [12;34]\Big\} \nonumber \\
&& +
{g_s^6 C_F \over N_c}\; \left( {s^2 + t^2 \over u^2}
-\frac{1}{N_c}\; {s^2\over  tu} \right)\;
\Big\{ 2C_F ([13] + [24]) + \frac{1}{N_c} [12;34]\Big\} \nonumber \\
& & -  {g_s^6 C_F \over N_c}\;  {s^2\over  tu}\;
 \left(1-\frac{1}{N_c^2}\right)
\Big\{  [12;34]\Big\} \nonumber \\
&& + \frac{g_s^4 g'^2}{N_c}\; 2\Re \left({ (s^2 + u^2)\{v^2 \}
+(s^2-u^2)\{a^2  \} \over t [ t-M_{Z'}^2
+  i M_{Z'} \Gamma_{Z'} ] } \right) \nonumber \\
&& \times
\; \Big\{ 2C_F ([14] + [23] - [12]-[34]) \Big\}   \nonumber \\
&& + \frac{g_s^4 g'^2}{N_c}\; 2\Re \left({ (s^2 + t^2)\{v^2 \}
+(s^2-t^2)\{a^2  \} \over u [ u-M_{Z'}^2
+ i M_{Z'} \Gamma_{Z'} ] } \right) \nonumber \\
&& \times
\; \Big\{ 2 C_F ([13] + [24] - [12]-[34]) \Big\}   \nonumber \\
&& + g_s^4 g'^2\frac{C_F}{N_c}\; 2\Re \left({2 s^2 \{v^2+a^2 \}
  \over t [ u-M_{Z'}^2
+ i M_{Z'} \Gamma_{Z'} ] } \right) \nonumber \\
&& \times
\; \Big\{ 2 C_F ([23] + [14])
+ \frac{1}{N_c}( [12] + [34]- [13]-[24] )  \Big\}   \nonumber \\
&& + g_s^4 g'^2\frac{C_F}{N_c}\; 2\Re \left({2 s^2 \{v^2+a^2 \}
  \over u[ t-M_{Z'}^2
+ i M_{Z'} \Gamma_{Z'} ] } \right) \nonumber \\
&& \times
\; \Big\{ 2 C_F ([13] + [24])
+ \frac{1}{N_c}( [12] + [34]- [23]-[14] )  \Big\}   \nonumber \\
&& + 2g_s^2 g'^4\; \left({ (s^2 + u^2)\{(v^2+a^2)^2 \}
+(s^2-u^2)\{4 v^2a^2 \} \over (t-M_{Z'}^2)^2
+ M_{Z'}^2 \Gamma_{Z'}^2} \right) \nonumber \\
&&\times
\; \Big\{ 2C_F ([13] + [24]) \Big\} \nonumber \\
&& + 2g_s^2 g'^4\; \left({ (s^2 + t^2)\{(v^2+a^2)^2 \}
+(s^2-t^2)\{4 v^2a^2 \} \over (u-M_{Z'}^2)^2
+ M_{Z'}^2 \Gamma_{Z'}^2} \right) \nonumber \\
&&\times
\; \Big\{ 2C_F ([23] + [14]) \Big\} \nonumber \\
&& + \frac{g_s^2 g'^4}{N_c}\; 2\Re \left({ s^2 \{(v+a)^4
+(v-a)^4 \}
\over [u-M_{Z'}^2 + i M_{Z'} \Gamma_{Z'}][t-M_{Z'}^2
- i M_{Z'} \Gamma_{Z'}]   } \right) \nonumber \\
&&\times
\; \Big\{ 2 C_F ([13] + [14]+ [23] + [24] - [12] - [34]) \Big\} \nonumber \\
\end{eqnarray}

\subsubsection*{$q(p_1)+\bar q(p_2) \to g(p_3)+ g(p_4) + g(k)$}
\begin{eqnarray}
\overline{\sum}\vert{\cal M}\vert^2 &= &
{g_s^6 C_F}\;(t^2 + u^2)\; \left[ \left(1-\frac{1}{N_c^2}\right)
\frac{1}{tu} - \frac{2}{s^2}  \right]\;
\Big\{ 2C_F [12] + 2N_c[34]\Big\} \nonumber \\
&& -  {g_s^6 N_c \over 4}\;(t^2 + u^2)\; \left[ \left(1-\frac{2}{N_c^2}\right)
\frac{1}{tu} - \frac{2}{s^2}  \right]\;
\Big\{ 2C_F [12;34]\Big\} \nonumber \\
&& +  {g_s^6 N_c \over 4}\;(t^2 - u^2)\; \left[
\frac{1}{tu} - \frac{2}{s^2}  \right]\;
\Big\{ 2C_F( [14] + [23]  - [13] - [24] ) \Big\} \nonumber \\
\end{eqnarray}

\subsubsection*{$q(p_1)+g(p_2) \to q(p_3)+ g(p_4) + g(k)$}
\begin{eqnarray}
\overline{\sum}\vert{\cal M}\vert^2 &= &
\frac{g_s^6}{2}\;(s^2 + u^2)\; \left[ \left(1-\frac{1}{N_c^2}\right)
\frac{1}{-su} + \frac{2}{t^2}  \right]\;
\Big\{ 2C_F [13] + 2N_c[24]\Big\} \nonumber \\
&& -  {g_s^6 N_c^2 \over 4(N_c^2-1)}\;(s^2 + u^2)\;
 \left[ \left(1-\frac{2}{N_c^2}\right)
\frac{1}{-su} + \frac{2}{t^2}  \right]\;
\Big\{ 2C_F [13;24]\Big\} \nonumber \\
&& +  {g_s^6 N_c^2 \over 4(N_c^2-1)}\;(s^2 - u^2)\; \left[
\frac{1}{-su} + \frac{2}{t^2}  \right]\;
\Big\{ 2C_F( [14] + [23]  - [12] - [34] ) \Big\}  \nonumber \\
\end{eqnarray}

\subsubsection*{$g(p_1)+g(p_2) \to g(p_3)+ g(p_4) + g(k)$}
\begin{eqnarray}
\overline{\sum}\vert{\cal M}\vert^2 &= &
\frac{4g_s^6N_c^2}{N_c^2-1}\;
\left[ 3 - \frac{ut}{s^2} - \frac{us}{t^2} - \frac{st}{u^2} \right]\;
\Big\{ \frac{2}{3}N_c ([12] + [34]+[13]+[14]+[23]+[24])\Big\} \nonumber \\
&&
+ \frac{2g_s^6N_c^2}{3(N_c^2-1)}\;
\left[ 3 + \frac{st}{u^2} + \frac{us}{t^2} -2 \frac{ut}{s^2} -3 \frac{s^2}{ut} 
\right]\;\Big\{ N_c ([12] + [34])\Big\} \nonumber \\
&&
+ \frac{2g_s^6N_c^2}{3(N_c^2-1)}\;
\left[ 3 + \frac{st}{u^2} + \frac{ut}{s^2} -2 \frac{us}{t^2} -3 \frac{t^2}{us} 
\right]\;\Big\{ N_c ([13] + [24])\Big\} \nonumber \\
&&
+ \frac{2g_s^6N_c^2}{3(N_c^2-1)}\;
\left[ 3 + \frac{ut}{s^2} + \frac{us}{t^2} -2 \frac{st}{u^2} -3 \frac{u^2}{st} 
\right]\;\Big\{ N_c ([14] + [23])\Big\}       \nonumber \\
\end{eqnarray}

Several comments are in order concerning the colour-suppressed interference
contributions which appear in the above results.
We note first that they are not singular when the soft  gluon is collinear 
with either the initial- or final-state partons.
Secondly, only two of them are actually independent.
To see this we introduce the following notation for the ``dipole" 
combinations:
\begin{eqnarray}
I_a & =&  [13]+[24]-[14]-[23] \; ,\nonumber \\
I_b & =& [14]+[23]-[12]-[34] \; ,\nonumber \\ 
I_c &= & [13]+[24]-[12]-[34]\; .
\end{eqnarray}
Then it follows that  (see  Ref.~\cite{DKT})
\begin{eqnarray}
I_a+I_b & = & I_c \; ,\nonumber \\
I_a +I_c & = & [ 13 ; 24 ] \; ,\nonumber \\           
I_b -I_a & = & [14 ; 23] \; ,\nonumber \\
{}[ 13 ; 24 ]+ [ 14 ; 23 ] & = & - [12;34]\; .
\end{eqnarray}
Note that interchanging  $ p_3 \leftrightarrow p_4$ gives
\begin{eqnarray}
I_a  &\to &  -I_a    \; ,\nonumber \\
I_b  &\to & I_c     \; ,\nonumber \\
{}[ 12 ; 34 ] &\to & [ 12 ; 34 ] \; ,\nonumber \\  
{} [ 13 ; 24 ] &\to & [ 14 ; 23 ]\; .
\end{eqnarray}

\end{document}